\documentclass[aps,prd,twocolumn,groupedaddress,showpacs,preprintnumbers,draft]
{revtex4}
\usepackage{epsf} 
\begin{document}
\preprint{BROWN-HET-14-04}
\preprint{MCGILL-16-04}
\def\Box{\nabla^2}  
\def\ie{{\em i.e.\/}}  
\def\eg{{\em e.g.\/}}  
\def\etc{{\em etc.\/}}  
\def\etal{{\em et al.\/}}  
\def\S{{\mathcal S}}  
\def\I{{\mathcal I}}  
\def\mL{{\mathcal L}}  
\def\H{{\mathcal H}}  
\def\M{{\mathcal M}}  
\def\N{{\mathcal N}} 
\def\O{{\mathcal O}} 
\def\cP{{\mathcal P}} 
\def\R{{\mathcal R}}  
\def\K{{\mathcal K}}  
\def\W{{\mathcal W}} 
\def\mM{{\mathcal M}} 
\def\mJ{{\mathcal J}} 
\def\mP{{\mathbf P}} 
\def\mT{{\mathbf T}} 
\def\mR{{\mathbf R}}
\def\mS{{\mathbf S}}
\def\mX{{\mathbf X}}
\def\mZ{{\mathbf Z}}
\def\eff{{\mathrm{eff}}}  
\def\Newton{{\mathrm{Newton}}}  
\def\bulk{{\mathrm{bulk}}}  
\def\brane{{\mathrm{brane}}}  
\def\matter{{\mathrm{matter}}}  
\def\tr{{\mathrm{tr}}}  
\def\nr{{\mathrm{normal}}}  
\def\implies{\Rightarrow}  
\def\half{{1\over2}}  
\newcommand{\da}{\dot{a}}
\newcommand{\db}{\dot{b}}
\newcommand{\dn}{\dot{n}}
\newcommand{\dda}{\ddot{a}}
\newcommand{\ddb}{\ddot{b}}
\newcommand{\ddn}{\ddot{n}}
\newcommand{\ba}{\begin{array}}
\newcommand{\ea}{\end{array}}
\def\be{\begin{equation}}
\def\ee{\end{equation}}
\def\bea{\begin{eqnarray}}
\def\eea{\end{eqnarray}}
\def\bs{\begin{subequations}}
\def\es{\end{subequations}}
\def\g{\gamma}
\def\G{\Gamma}
\def\vp{\varphi}
\def\mpl{M_{\rm P}}
\def\ms{M_{\rm s}}
\def\ls{\ell_{\rm s}}
\def\lp{\ell_{\rm pl}}
\def\l{\lambda}
\def\gs{g_{\rm s}}
\def\d{\partial}
\def\co{{\cal O}}
\def\sp{\;\;\;,\;\;\;}
\def\spa{\;\;\;}
\def\r{\rho}
\def\dr{\dot r}
\def\dt{\dot\varphi}
\def\e{\epsilon}
\def\k{\kappa}
\def\m{\mu}
\def\n{\nu}
\def\om{\omega}
\def\tn{\tilde \nu}
\def\p{\phi}
\def\vp{\varphi}
\def\P{\Phi}
\def\r{\rho}
\def\s{\sigma}
\def\t{\tau}
\def\x{\chi}
\def\z{\zeta}
\def\a{\alpha}
\def\b{\beta}
\def\de{\delta}
\def\bra#1{\left\langle #1\right|}
\def\ket#1{\left| #1\right\rangle}
\newcommand{\stt}{\small\tt}
\renewcommand{\theequation}{\arabic{section}.\arabic{equation}}
\newcommand{\eq}[1]{equation~(\ref{#1})}
\newcommand{\eqs}[2]{equations~(\ref{#1}) and~(\ref{#2})}
\newcommand{\eqto}[2]{equations~(\ref{#1}) to~(\ref{#2})}
\newcommand{\fig}[1]{Fig.~(\ref{#1})}
\newcommand{\figs}[2]{Figs.~(\ref{#1}) and~(\ref{#2})}
\newcommand{\GeV}{\mbox{GeV}}
\def\ricci{R_{\m\n} R^{\m\n}}
\def\riemann{R_{\m\n\l\s} R^{\m\n\l\s}}
\def\triemann{\tilde R_{\m\n\l\s} \tilde R^{\m\n\l\s}}
\def\tricci{\tilde R_{\m\n} \tilde R^{\m\n}}
\title{Single Field Baryogenesis}
\author{K.R.S. Balaji $^{1}$ \email[Email:]{balaji@hep.physics.mcgill.ca}
and Robert H. Brandenberger $^{2,3,1}$ \email[Email:]{rhb@het.brown.edu}}
\affiliation{1) Department of Physics, McGill University, Montr\'eal, QC, 
Canada H3A 2T8}
\affiliation{2) Department of Physics, Brown University, Providence, RI 02912, 
USA} 
\affiliation{3) Perimeter Institute for Theoretical Physics,
35 King Street N., Waterloo, ON N2J 2W9, CANADA.}
\begin{abstract}
We propose a new variant of the Affleck-Dine baryogenesis mechanism in which
a rolling scalar field couples directly to left- and right-handed neutrinos,
generating a Dirac mass term through neutrino Yukawa interactions. 
In this setup, there are no explicitly $CP$
violating couplings in the Lagrangian. The rolling scalar field is also
taken to be uncharged under the $B - L$ quantum numbers. During the
phase of rolling, scalar field decays generate a non-vanishing number
density of left-handed neutrinos, which then induce a net baryon number
density via electroweak sphaleron transitions.
\end{abstract}
\maketitle

\section{Introduction}
The origin of the observed baryon to entropy ratio remains an
unsolved problem. There are many scenarios which provide
ways of explaining the origin of the asymmetry between
baryons and antibaryons
(for recent reviews, see e.g. \cite{Dolgov,Trodden,RubShap}).

As realized a long time ago by Sakharov \cite{Sakharov}, in order to
generate a baryon number asymmetry from symmetric initial conditions,
three criteria need to be satisfied. There need to be baryon number
violating processes, there must be C and CP violation, and the
processes which produce the baryon asymmetry should take place out
of thermal equilibrium. 

The first models of baryogenesis were based on the out-of-equilibrium
decay of super-heavy Higgs and gauge particles in grand-unified theories (GUT)
\cite{GUTBG}. However, in the context of inflationary cosmology \cite{Guth}
(the predictions of which - namely spatial flatness of the Universe
and an almost scale-invariant spectrum of adiabatic density fluctuations -
are well supported by the most recent observational results \cite{WMAP}),
it is likely that, after inflation, the universe will not reheat to
a temperature sufficiently high to produce these super-heavy GUT particles.

In the light of inflationary cosmology, another paradigm for baryogenesis,
namely the Affleck-Dine (AD) scenario \cite{AD}, has attracted an increasing
amount of attention. Crucial to the AD scenario is the rolling of a scalar
field $\phi$ (belonging to the sector of particle
physics beyond the ``Standard Model'') 
which couples to standard model fields and whose decay produces the
observed asymmetry. It is natural to assume that this scalar field has been displaced 
from its low temperature ground state during the period of inflation,
either by quantum vacuum fluctuations or by initial classical perturbations. 
It is usually assumed that this field carries a non-vanishing $B - L$ quantum
number, such that the decay generates a baryon asymmetry. Since the scalar
field dynamics singles out a preferred direction in field space, $CP$ is
violated dynamically (note that the Lagrangian is symmetric under $CPT$ but
the cosmological background dynamics sets a preferred time direction).
Also, the dynamics of $\phi$ is an out-of-equilibrium process, and thus, it is clear
that all of the Sakharov conditions are satisfied.

As mentioned above, within the AD baryogenesis scheme, the rolling 
complex field is subjected to Hubble friction which then produces a $CP$ violating charge 
asymmetry which subsequently is transformed to a
baryon asymmetry. There are other possible variations to this basic setup and they 
broadly fall in one of the following types:
(i) the field $\phi$ has scalar-gauge interactions which are $CP$ and 
$CPT$ violating like in spontaneous baryogenesis \cite{Kaplan},  
or (ii) $\phi$ has $CP$ violating Yukawa interactions with leptons, and 
the final baryon asymmetry arises via sphaleron transitions from the  
lepton asymmetry produced in the first stage, as in the model of non-thermal 
leptogenesis \cite{Shafi}. Generically, these models
involve an interaction Lagrangian of the form $\kappa O_\mu j^\mu_{B/L}$, 
where $j^\mu_{B/L}$ denotes a baryon and/or lepton number violating current, $O_{\mu}$ is
an operator vector field depending on the rolling scalar field $\phi$, 
and $\kappa$ is a coupling constant which could in principle be complex
(thus introducing new explicit $C$- and $CP$-violating couplings into the Lagrangian).  
The form of $O_\mu$ and the specific choice of the current $j^{\mu}$ depend
on the specific variation. For instance, in the case of spontaneous 
baryogenesis \cite{Kaplan}, $O_\mu \sim \partial_\mu \phi$. For the
recently proposed gravitational baryogenesis \cite{gbaryo} scenario,
the role of the scale field $\phi$ is played by the Ricci scalar $R$ and 
$O_\mu \sim \partial_\mu R$. In both of these examples, the current 
$j^{\mu}$ is the $B - L$ current.
In the case of non-thermal leptogenesis, $j^\mu_L$ is the lepton current
which couples to the scalar current $O_\mu \sim \phi_\mu$ and breaks the
$B-L$ quantum number by two units, thereby generating a Majorana mass term for
neutrinos. All of the above proposals require
(i) presence of a baryon or a lepton charge for the 
rolling scalar field and/or (ii) $CP$ violation in the interaction Lagrangian 
via complex couplings.
 
Here, we wish to present a new variant of AD baryogenesis which neither
requires the rolling scalar field to carry non-vanishing baryon or lepton
charges, nor does
it involve $CP$-violating couplings in the Lagrangian. Analogously to
what happens in leptogenesis \cite{lepto} (another currently popular route
to explaining the observed baryon to entropy ratio), in an initial step
the scalar field $\phi$ first decays into neutrinos, producing an asymmetry
in the left-handed leptons which is then converted, making use of the
sphaleron transitions \cite{sphaleron} of the standard electroweak theory,
into a net baryon charge density. Note that total lepton number is not
violated: the asymmetry in left-handed neutrinos is compensated by a
corresponding asymmetry in the right-handed neutrino sector.
In contrast to usual leptogenesis
scenarios, in our case the neutrinos are of Dirac type
\footnote{See also \cite{Lindner} for another baryogenesis model which
makes use of Dirac neutrinos. However, in that work, the scalar fields
decay into purely right-handed neutrinos, instead of inducing
Dirac mass terms as they do in our work. In terms of using time-dependent
fermion mass terms, our scenario has analogies with ``coherent
baryogenesis'' \cite{coherent}.}. Note that there are some similarities between
our scenario and the {\it electrogenesis} proposal of \cite{Tomislav}
which also involves the decay of a scalar condensate into leptons.

In our scenario, the Sakharov conditions are satisfied in the following way:
the baryon number violating processes are electroweak sphalerons which 
convert an asymmetry in left-handed lepton number to baryon number. The
asymmetry in the left-handed leptons is generated by Yukawa neutrino
interactions of $\phi$, which generate a time-dependent effective Dirac
mass for the neutrinos. Total lepton number is not violated explicitly. Thus,
there is an asymmetry in the right-handed neutrinos which compensates
for the asymmetry in the left-handed neutrinos. Neutrino interactions
are too weak to equilibrate right- and left-handed neutrinos (for other
leptons in the Standard Model, the equilibration would be rapid). The
$CP$ symmetry is violated dynamically via the asymmetric initial conditions
of $\phi$ (more specifically, the complex phase of $\phi$) in our Hubble
patch set up by a phase of primordial inflation. In this respect, our
work is related to recent work on dynamical $CP$ violation in the
early universe \cite{BBBL} (developing earlier ideas of Dolgov \cite{Dolgov}).
Finally, since the
dynamics of $\phi$ is out of thermal equilibrium, the third Sakharov
criterion is trivially satisfied.

\section{Generation of the Lepton Number Asymmetry}

Our starting point is a Yukawa-type interaction Lagrangian 
$L_Y$ which couples the scalar
field $\phi$ to left and right handed neutrinos $\nu_L$ and $\nu_R$: 
\be \label{intlag}
L_Y \, = \, y_\nu \bar\nu_L \nu_R \phi + h.c.~.
\label{yukawa}
\ee
In order that this Lagrangian be a $SU(2)_L$ scalar, the field
$\phi$ needs to possess $SU(2)_L$ quantum numbers. Thus, $\phi$ is
unlikely to be the inflaton (the field driving inflation), because
a gauge non-singlet field generically obtains a potential which is too steep
to allow for a phase during which it rolls sufficiently slowly to yield 
inflation \footnote{In models which obtain small Dirac couplings by the 
Frogatt-Nielson \cite{FN} technique, our rolling scalar could be the 
singlet inflaton field.}. We take the couplings $y_\nu$ to be real
numbers. Thus, we are not introducing any new $CP$-violating
phases into the Lagrangian, and the Lagrangian does not
contain any terms which violate baryon or lepton number. 
The interaction Lagrangian (\ref{intlag}) 
generates Dirac neutrino masses, and we will see below that it
allows the decay of a $\phi$ condensate displaced from the ground state
to produce an asymmetry in the left- and right-handed lepton numbers
(with the total lepton number being conserved).

In the context of theories beyond the standard model of particle physics,
there are many possible candidates for $\phi$, e.g. the scalar partner
of one of the standard model fermions, in particular the sneutrino (used
for ordinary leptogenesis in \cite{sneutrino}). Such condensates are
excited during inflation by quantum fluctuations, and due to the 
exponential red-shifting of length scales during inflation, the condensates
will then be quasi-homogeneous on scales larger than the Hubble radius after
the end of inflation. The condensates are frozen until the Hubble constant
decreases to a value comparable to the condensate mass, after which the
field will begin to roll towards it's ground state.

Let us first examine the dynamics of a complex scalar field condensate $\phi$, 
which in the slow-roll approximation (this approximation is not
crucial for the mechanism, but is chosen to make the analysis specific)
is described by
\be
3H\dot\phi + V^\prime \, = \, 0 \, .
\label{rolleqn}
\ee
If the phase of $\phi$ is non-vanishing \footnote{Note that the phase
of $\phi$ cannot be rotated away without introducing $CP$-violating
phases in the interaction Lagrangian. As also discussed in \cite{BBBL},
having a non-vanishing initial phase of $\phi$ after inflation is completely
natural in the context of early universe (inflationary) cosmology.}, 
then the dynamical evolution
of $\phi$ (the motion which breaks the time-translational symmetry) may lead
to a charge asymmetry density (see e.g. \cite{Trodden}) for the decay
particles
\be
Q_\phi \, =  \, -\frac{Im(\phi^* dV / d\phi^{*})}{3H} \, .
\label{chrasym}
\ee
This equation is obtained from the equation
of covariant conservation of the scalar current,
evaluated in the slow roll approximation. If the
potential is a function of $\phi^* \phi$, then $Q_{\phi}$
vanishes. However, if for example it is a function of
$\phi^2$ plus complex conjugates, then it does, in general,
not vanish. 

Given the Yukawa interaction Lagrangian (\ref{intlag}), the 
decay process being considered is $\phi  \to \bar\nu_R\nu_L$. The
perturbative decay rate is
\be
\Gamma_\phi \, = \, \frac{y_\nu^2m_\phi}{8 \pi} \, .
\label{decay}
\ee
Since $Q_\phi \neq 0$, the changes in the number densities 
$n(\phi)$ and $n(\phi^*)$ are not the same. As a result, the decay of the 
condensate produces an asymmetry in the number density of the decay leptons whose
value is given by
\be
A_\nu \, \equiv \, n_{\nu} - n_{\bar \nu} \, =  \, Q_\phi \frac{\Gamma_\phi}{m_\phi}~.
\label{lepdiff}
\ee

In the above, we are assuming that the total decay rate of $\phi$ into
particles is less important in the equation of motion of the condensate
than the decay due to Hubble damping (otherwise the use of (\ref{rolleqn})
would not be consistent. We focus on the decay of $\phi$ into neutrinos, since, as we
shall see below, any asymmetry in the right- and left-handed number densities
of other leptons will wash out. However, due to their very weak interactions,
the washout for neutrinos is ineffective.

The lepton to entropy ratio resulting from (\ref{lepdiff}) is
\be
N_\nu \, \equiv \, \frac{A_\nu}{n_\gamma} \, = \,
\frac{Q_\phi y_\nu^2}{8\pi n_\gamma}~.
\label{lepden}
\ee
Note that although, $Q_\phi$ is $CPT$ violating, $N_\nu$ is not (since
$n_\gamma$ is $CPT$ odd) and hence can be affected by reverse equilibrating 
processes. In the following, we track the left-handed component $N_L$ of the 
asymmetry in (\ref{lepden}) 
during the subsequent evolutionary course of the Universe.

\section{Study of the Equilibration}

Recall that while the decay of $\phi$ does not violate total lepton number,
it generates an asymmetry in the number 
densities of left- and right-handed Dirac neutrinos:
\be
n(\bar\nu_L) \neq n(\nu_L) ~; ~ n(\bar\nu_R) \neq n(\nu_R) ~.
\label{nudiff}
\ee
The electroweak sphalerons then convert part of the asymmetry in the
left-handed neutrinos into a net baryon asymmetry. However, a necessary 
condition for this scenario to work is that the asymmetry 
in (\ref{nudiff}) does not equilibrate during the time interval
between $\phi$ decay and electroweak symmetry breaking. In the
following, we will show that because of their small masses and coupling
constant, the neutrino number densities do not equilibrate, whereas the
number densities of other leptons in the Standard Model would. 

The net left-handed neutrino to entropy ratio $N_L$ in a homogeneous and
isotropic universe evolves according to the kinetic equation \cite{luty}
\be
\frac{dN_L}{dz}\propto \frac{dY_{\nu_L}}{dz} = - 
\frac{z}{sH(m_\phi)}\Big (\frac{Y_{\nu_L}}{Y_{\nu_L}^{eq}}\Big )
\gamma(\nu_L \to \nu_R) \, ,
\label{kinetic}
\ee
where $M_p$ is the Planck mass, $z$ is proportional to the cosmological
redshift via the relation $T = m_\phi/z$, where $T$ is the temperature of
radiation, $Y_X$ denotes the ratio between the number density of $X$ particles
and the entropy density (and the superscript {\it eq} indicates the
value of the corresponding quantity in thermal equilibrium), and $\gamma$
is proportional to the cross section (see below). Note that the effects
of the Hubble expansion cancel out if we track the ratio of particle
number densities. Note also that we are neglecting the production of
neutrinos by back-reaction effects.

We note a caveat in using (\ref{kinetic}): it assumes a thermal 
distribution of particles, and since the $\phi$ condensate is not a thermal
distribution of $phi$ particles, the decay products will not be thermally 
distributed, either. However, the purpose of the following discussion is 
to get an order of magnitude estimate of the washout, and for this
purpose, the use of (\ref{kinetic}) should be adequate.

Assuming that we are working in the radiation dominated era, we have
\be
Y_i = \frac{n_i}{s}=\frac{45 n_i}{2g_* \pi^2 T^3}~;~Y_i^{eq} =
\frac{45 g_i m_i^2 K_2(z)}{4 \pi^4 g_* T^2}~ \, ,
\label{defns}
\ee
where $K_i$ are the modified Bessel functions. 
If the reaction rates are assumed to be dominated by decay
processes with width $\Gamma$ (see below for more details), then \cite{twolf}
\be
\gamma = n_i^{eq} \frac{K_1(z)}{K_2(z)} \Gamma_i 
\equiv n_i^{eq} \kappa (z) \Gamma_i =  \frac{g_i T m_i^2K_1(z)}{2\pi^2} \Gamma_i \, ,
\label{gamma}
\ee
where $\gamma_i$ has mass dimension of 4..

Within the standard model, there are three different contributions to $\gamma$. 
These are: (i) due to mass, (ii) inverse scattering of $\nu_L$ leading to resonant
production of $Z$ and $W^\pm$ gauge bosons and (iii) due to neutral and 
charged current scattering processes.

Given the sizeable decay widths of the gauge bosons to leptonic final states 
($\Gamma_W \sim 0.21 ~\mbox{GeV}~,~\Gamma_Z \sim 0.17 ~\mbox{GeV}$) any 
production of these states via an inverse scattering of left-handed neutrinos 
will quickly replenish the left-handed states. 
Neutral and charged scattering processes will also not deplete the 
number density of left-handed leptons. The subsequent conversion of these final 
states to right-handed states (again via mass terms) are strongly suppressed and can 
be neglected to leading order \cite{maalampi}. All of these arguments hold as long 
as the Lagrangian conserves lepton number as it is in our case. 

As a result, we only need to consider the depletion due to a small neutrino mass. 
Thus,
\be
\gamma \equiv \gamma ^m = \Big(\frac{m_\nu}{T}\Big )^2 \Gamma_\phi \kappa(z) 
Y_{\nu_L}^{eq} s~,
\label{lrrate}
\ee
which is helicity suppressed at high temperatures. The presence of
$\Gamma_\phi$ denotes the production rate of $\nu_L$ states. 

In order to get a more quantitative feeling for the loss of left-handed neutrino 
due to conversion, let us examine $\gamma^m$ in the asymptotic limit of large 
$z$ which corresponds to late times. In this case, $\kappa(z) \approx 1$
and therefore,
\be
\gamma = \gamma^m \approx \Big(\frac{m_\nu}{T}\Big )^2 \Gamma_\phi Y_{\nu_L}^{eq}s~.
\label{largekappa}
\ee
Substituting (\ref{largekappa}) into (\ref{kinetic}) we have
\be
\frac{dY_{\nu_L}}{dz} \approx - 
\frac{zY_{\nu_L} \Gamma_\phi}{H(m_\phi)}\Big(\frac{m_\nu}{T}\Big )^2 ~.
\label{approxrate}
\ee
As a function of temperature,
\bea
Y_{\nu_L}(T) &=& Y_{\nu_L}^0 \exp 
\Big(\frac{2}{H(m_\phi)}\int dT \frac{m_\phi^2 m_\nu^2\Gamma_\phi}{T^5} \Big )~, 
\nonumber\\
&=& Y_{\nu_L}^0 \exp
\Big(-\frac{m_\phi^2 m_\nu^2\Gamma_\phi}{2T^4H(m_\phi) } \Big )~.
\label{ynul}
\eea
The initial value $Y_{\nu_L}^0$ is proportional to the number density of 
$\phi$ particles divided by the entropy, evaluated at the time when the $\phi$
condensate decays (see Section II). 

Clearly, the equilibration process is negligible, i.e. 
$Y_{\nu_L} \simeq Y_{\nu_L}^0$, if
\be
\frac{m_\phi^2 m_\nu^2\Gamma_\phi}{2T^4 H(m_\phi) } \ll 1~.
\label{cond}
\ee
Conversely, if the left hand side exceeds 1, then most of the initial asymmetry
disappears. From the form of (\ref{cond}) it is clear that the equilibration
process is most efficient at the lowest temperature being considered. 

In our context, the lowest temperature is the scale of electroweak symmetry 
breaking. If the neutrino asymmetry survives until that time, it gets converted 
into a baryon asymmetry by electroweak sphaleron effects. Thus, in evaluating
(\ref{cond}) for neutrino equilibration, we set
$T\geq 1$TeV, use as neutrino mass the value
$m_\nu \sim 0.01$ eV as suggested by current data, and thus conclude
that (\ref{cond}) is satisfied as long as $\Gamma_\phi$ is not too much
larger than $T$. Hence, any asymmetry in the left-handed neutrino density
established via decay of the $\phi$ condensate at an early stage survives
until the scale of electroweak symmetry breaking. For other leptons, this is
not the case.

\section{Discussion and Conclusions}

We have presented a variation of the AD baryogenesis scenario in which neither
new baryon number violating processes nor new sources of $CP$-violation are
introduced. As in the conventional AD setup, a scalar field condensate,
displaced from its ground state during a period of primordial inflation,
produces a net asymmetry in left-handed neutrinos (compensated by a
corresponding asymmetry in the right-handed neutrinos) during the time
interval when it is rolling towards its ground state in the post-inflationary
period. 

We have established that an initial asymmetry in the density of left-handed
neutrinos will survive until the scale of electroweak symmetry breaking. At
this point, it will be converted to a 
baryon asymmetry via sphaleron transition. The resulting baryon to entropy
ratio $A_B$ is related to the left-handed neutrino asymmetry via
\be
A_B \, = \, \alpha A_\nu \, ,
\label{basym}
\ee
where the $\alpha$ is a constant of order unity whose precise value depends on
the number of generations and on the number of Higgs particles.

{\bf Acknowledgments}:

We thank Guy Moore for several clarifying discussions and
Tomislav Prokopec for very useful comments on the draft of this
paper.. RB is supported 
in part by the US Department of Energy under Contract DE-FG02-91ER40688, TASK~A. 
He thanks the Perimeter Institute for hospitality and
financial support. The work of KB is supported by NSERC (Canada) and by the Fonds 
de Recherche sur la Nature et les Technologies du Qu\'ebec.

\end{document}